\documentclass{ws-ijmpa}
\usepackage{amsmath}
\def\p{\partial}

\def\g{\gamma}

\def\d{\delta}
\def\de{\delta}
\def\D{\Delta}
\def\De{\Delta}
\def\ld{\lambda}

\def\Ld{\Lambda}
\def\ep{\epsilon}
\def\e{\eta}

\def\om{\omega}
\def\rh{\rho}
\def\s{\sigma}

\def\b{\beta}

\def\a{\alpha}
\def\ast{\alpha^*}

\def\pdellx'{\frac{\partial}{\partial x'}}
\def\pdellw'{\frac{\partial}{\partial w'}}
\newcommand{\be}{\begin{equation}}
\newcommand{\ee}{\end{equation}}
\def\bed{\begin{displaymath}}
\def\eed{\end{displaymath}}
\def\bea{\begin{eqnarray}}
\def\eea{\end{eqncrray}}
\def\[{$$}
\def\]{$$}
\newcommand{\beas}{\begin{eqnarray*}}
\newcommand{\eeas}{\end{eqnarray*}}

\newcounter{stokes}

\begin{document}
%\maketitle
\title{Violation of $U_1$ Gauge Symmetry by Yang-Mills Gravity and Deflection of Light Experiment }
\author{
Leonardo Hsu\\
Department of Chemistry and Physics,   Santa Rosa Junior College\\
Santa Rosa, CA 95401, USA\\
Jong-Ping Hsu and Yun Hao\\%\footnote{E-mail: jhsu@umassd.edu}\\
 Department of Physics, 
 University of Massachusetts Dartmouth\\ 
North Dartmouth, MA 02747, USA\\
} 
%\maketitle
\date{April 2018}
\maketitle
{\small   Based on the gauge symmetry framework, the $U_1$ symmetry of electrodynamics is violated in the presence of gravity with space-time translational gauge symmetry in inertial frames.  For a light ray, an eikonal equation with effective metric tensors is derived in the geometric-optics limit.  Under these conditions, the angle of the deflection of light by the sun is calculated to be $\d \phi \approx  1.75''$ in inertial frames without requiring a gauge condition such as $\p_\mu A^\mu=0$.  In contrast, if the theory is $U_1$ gauge invariant, one can impose the gauge condition $\p_\mu A^\mu=0$ in the derivation of the eikonal equation.  In this case, one obtains a slightly different effective metric tensor and a different angle of deflection $\d \phi \approx  1.52''$.   However, because the precision of experiments in the last century using optical frequencies has been no better than (10 $-$ 20)\% due to large systematic errors, one cannot unambiguously rule out the result  $\d \phi \approx  1.52'$.  It is hoped that the precision of these data can be improved in order to test Yang-Mills gravity.}

\bigskip



\bigskip
\section{Introduction}
The formulations of
electromagnetic and general  gauge theories associated with internal gauge
groups are all based on the replacement,
$\p_\mu \to \p_\mu-ifB_\mu$ in the Lagrangian.  The field $B_\mu=B_\mu^a t_a$
involves constant matrix representations of
the generators $t_a$ of the groups associated with internal gauge symmetry.  However, Yang-Mills gravity is based on the external space-time translation gauge group $T_4$.
Since the generators of the group $T_4$ are the displacement operators, $p_\mu=i\p _\mu$ \
(in natural units $c$=$\hbar$=1), the replacement in the Lagrangian is given by\cite{1,2,3}
\be
 \p _\mu
\to \p _\mu  -ig\phi_{\mu\nu} p^\nu \equiv J_{\mu\nu}\p^\nu, \ \ \ \  
\ee
%%%%%%%%%%%%%%%%%%%%%%%%%1
$$
J_{\mu\nu}=\e_{\mu\nu}+g\phi_{\mu\nu}, \ \ \ \  \e_{\mu\nu}=(1,-1,-1,-1),
$$
 in inertial frames, where $g$ is the coupling constant associated with the gravitational tensor gauge field $\phi_{\mu\nu}$. 
 
 For readers not familiar with Yang-Mills gravity, we will briefly explain the basic ideas and the  Lagrangian with $T_4$ gauge symmetry.   Quantum Yang-Mills gravity is based the spacetime translational group ($T_4$) and involves (Lorentz) vector gauge functions $\Ld^\mu(x)$ and the related Hamilton's characteristic phase function in  inertial frames.\cite{1,2,3}  As a result, we can quantize the gravitational field without difficulty.  The $T_4$ gauge fields are associated with  the $T_4$ group and the generator $p_{\mu}=i \p_{\mu}$. The $T_4$ gauge covariant derivative  $J_{\mu\nu}\p^\nu= \p _\mu  -ig\phi_{\mu\nu} p^\nu$ is basic in gauge field theory of gravity and are given by the replacement (1), which dictates the universal interaction between gravitational and any other fields.\cite{1,2}

  As usual, the $T_4$ gauge curvature $C_{\mu\nu\a}$ is derived from the commutator of the $T_4$ gauge covariant derivative $J_{\mu}^{\ld}\p^\ld$, i.e.,
$
[J_{\mu}^{\ld}\p_{\ld}, J_{\nu}^{\s}\p_{\s}]=C_{\mu\nu\a}\p^\a, 
$
where $
 C_{\mu\nu\a}=J_{\mu\ld}(\p^\ld J_{\nu\a})-J_{\nu\ld}(\p^\ld J_{\mu\a}), \ \ \ \  J_{\mu\ld}=J_\mu^\b \e_{\b\ld}.$
The action $S_{\phi\psi}$ of Yang-Mills gravity for the tensor field $\phi_{\mu\nu}$ and a charged fermion field $\psi$ in an inertial frame is quadratic in the gauge curvature $C_{\mu\nu\a}$,
\be
S_{\phi\psi}=\int L_{\phi\psi} d^4 x, \ \ \ \   L_{\phi\psi}= L_{\phi} + L_{\psi},
\ee
%%%28%30%%4%%2 
\be   
L_{\phi}= \frac{1}{4g^2}\left (C_{\mu\nu\a}C^{\mu\nu\a}- 2C_{\mu\a}^{ \ \ \  \a}C^{\mu\b}_{ \ \ \  \b} \right),
\ee
%%%29%%31%%5%%3
\be
L_{\psi}=+ \overline{\psi}i\g^{\mu}(\p_\mu +g\phi_\mu^\nu \p_\nu -ie A_\mu) \psi  - m\overline{\psi}\psi.  
\ee
%%%%30%%31%%32%%6%%4
 The action $S_{\phi\psi}$ is invariant under local $T_4$ gauge transformations, although the Lagrangian density $L_{\phi}$ by itself is not invariant due to the presence of a total derivative term, which does not contribute to the gravitational field equations.\cite{3}  
 
 In quantum Yang-Mills gravity, the symmetric tensor field $\phi_{\mu\nu}$ is a massless spin-2 gauge boson.  The gravitational quadrupole radiation has been discussed with the usual gauge condition $\p^\mu \phi_{\mu\nu}=\p_\nu \phi^\ld_\ld /2$.  We have also calculated the power emitted per unit solid angle in the direction ${{\bf x}/|{\bf x}}|$ and that radiated by a body rotating around one of the principal axes of the ellipsoid of inertia.  The results to the second order approximation are the same as that obtained in general relativity and consistent with experiments.\cite{1,3} 

 In the geometric-optics limit,\cite{1,2} the fermion wave equation reduces to a Hamilton-Jacobi  type equation,
\be
G^{\mu\nu} (\p_\mu S)(\p_\nu S) -m^2=0, \ \ \  G^{\mu\nu}=\e_{\a\b}J^{\a\mu} J^{\b\nu}.
\ee
%%34%%8%%5
This equation of motion for macroscopic objects in flat space-time involves a new effective Riemannian metric tensor $G^{\mu\nu}$, which is actually a function of the $T_4$ gauge field $\phi_{\mu\nu}$ in Yang-Mills gravity.   It  is formally the same as the corresponding equation of motion for macroscopic objects in general relativity.\cite{3}\footnote{This equation obtained in the geometric-optics limit in Yang-Mills gravity based on inertial frames involves the effective metric tensor $G^{\mu\nu}$.  It is formally the same as the corresponding equation in general relativity. We call it the 'Einstein-Grossmann (EG) equation' in recognition of their collaboration.}  This equation is crucial for Yang-Mills gravity to be consistent with the perihelion shift of the Mercury, the deflection of light by the sun and the equivalence principle.\cite{1,3}
 
A satisfactory theory of gravity should be able to give a simple explanation why the gravitational force is attractive rather than repulsive.  Let us consider the  gravitational ($T_4$) tensor field $\phi_{\mu\nu}(x)$ and the electromagnetic potential field $A_\mu(x)$ in the gauge covariant derivative  and its complex conjugate in the fermion Lagrangian (4),
\be
 \p_\mu -ig\phi_\mu^\nu p_\nu -ie A_\mu +.... = \p_\mu +g\phi_\mu^\nu \p_\nu -ie A_\mu +....
\ee
%%%%%%08%%35%%9%%6
\be
 (\p_\mu -ig\phi_\mu^\nu p_\nu -ie A_\mu +....)^* = \p_\mu +g\phi_\mu^\nu \p_\nu + ie A_\mu +....
\ee
%%%%08%%9%%36%%10%%7
The gauge covariant derivative (6) and its complex conjugate (7) appear respectively in the wave equations of the electron (i.e., particle with charge $-e<0$) and the positron (i.e., antiparticle with charge $e>0$).  The electric force between two charged particles is due to the exchange of a virtual photon.  In quantum electrodynamics, this can be pictured in the Feynman diagrams with two vertices connected by a photon propagator.  The key properties of the electric force $F_e(e^-,e^-)$ (i.e., between electron and electron) and the force $F_e(e^-, e^+)$ (i.e., between electron and positron) are given by the third terms in (6) and in (7), i.e.,
$$
F_e(e^-,e^-): \ \ (-ie)\times (-ie)= - e^2, \ \ \ \ \   repulsive, 
$$
%%%09%%10%%
$$
F_e(e^-, e^+): \ \  (-ie)\times (ie)=+e^2,  \ \ \ \ \ \   attractive,
$$
and the force $F_e(e^+,e^+)$ is the same as $F_e(e^-,e^-)$. Thus, we have experimentally established attractive and repulsive electric forces, which are due to the presence of $i$ in the electromagnetic $U_1$ gauge covariant derivative $\p_\mu-ieA_\mu$.  The Yang-Mills gravitational force $F_{YMg}(e^-, e^-)$ (i.e., between electron and electron) and the force $F_{YMg}(e^-, e^+)$ (i.e., between electron and positron) are respectively given by the second terms in (6) and in (7).   Because the gravitation coupling terms in (6) and (7) do not involve $i$, we have only an attractive gravitational force,
$$
F_{YMg}(e^-, e^-) : \ \ \ (g) \times (g)= +g^2, \ \ \ \ \   attractive,  
$$
%%%0.10%%
$$
F_{YMg}(e^-, e^+) : \ \ \  (g) \times (g) =+g^2, \ \ \ \ \   attractive,
$$
and $F_{YMg}(e^+, e^+)$ is the same as $F_{YMg}(e^-, e^-)$.  Note that these qualitative  results for forces $F_e(e^-,e^-)$ and $F_{YMg}(e^-, e^-)$ are independent of the signs of the coupling constants $e$ and $g$.  
 
\section{The coupling of electromagnetic and gravitational fields with $T_4$ gauge symmetry}

  In the formulation of gauge field theory, once a gauge symmetry group is postulated, the coupling of the gauge field must be dictated by the gauge covariant derivative associated with the group.  In particular, the partial derivative $\p_\mu$ in the usual electromagnetic Lagrangian $(-1/4)\e^{\mu\a}\e^{\nu\b}(\p_\mu A_\nu - \p_\nu A_\mu)(\p_\a A_\b-\p_\b A_\a)$ must be modified according to (1).  Thus, in the presence of gravity with $T_4$ gauge symmetry, the electromagnetic Lagrangian $L_{em}$ in inertial frames must take the form
\be
L_{em}=-\frac{1}{4}\e^{\mu\a}\e^{\nu\b}F_{\mu\nu}F_{\a\b},   \ \ \ \  \e^{\mu\a}=(1,-1,-1,-1),
\ee
%%34%%%2%%8
$$  F_{\mu\nu}=\Delta_\mu A_{\nu}-\Delta_\nu A_{\mu}, \ \ \ \ \
\Delta_\mu=J_{\mu\nu}\p^{\nu}, 
$$
%$$%%%%%%
 according to the general principle of gauge symmetry. 

  Once the gravity is assumed to be a gauge field theory with local space-time translation ($T_4$) symmetry, the electromagnetic Lagrangian $L_{em}$ in (8) must have the $T_4$ gauge covariant derivative which dictate the presence of the term involving $g$, symmetric tensor field $\phi_{\mu\nu}$ and the $T_4$ generator $p_\mu$. This terms explains why the gravitational force is always attractive, in contrast to the electromagnetic force, as discussed in the introduction.  This term in the $T_4$ gauge covariant derivative dictates the $T_4$ gauge curvature $C_{\mu\nu\a}$, and the $T_4$ gauge invariant Lagrangian must be quadratic in the gauge curvature, as shown in (3).  All these properties, including the violation of the electromagnetic $U_1$ symmetry, are  consequence of the principle of gauge symmetry for the formulation of gauge field theories. 

%%%%%%%%%%%%%%%%%%%%%%%%%%%%%%%%
 One can demonstrate that $F_{\mu\nu}=\De_\mu A_\nu -\De_\nu A_\mu$ and the Lagrangian $L_{em}$  in (8) are not $U_1$ gauge invariant.
 Since the gravitational gauge field $\phi_{\mu\nu}$ is symmetric in $\mu$ and $\nu$, it is natural to have the following $U_1$ gauge transformations for the electromagnetic and gravitational gauge fields with an arbitrary and infinitesimal scalar function $\Ld(x)$,
\be
A_\mu(x) \to A'_\mu(x)=A_\mu(x) + \p_\mu \Ld (x), \ \ \ \ \ 
\ee
%%%%%3%%9
\be
\phi_{\mu\nu}(x) \to \phi'_{\mu\nu}(x)=\phi_{\mu\nu}(x) +\p_\mu \p_\nu \Ld(x).
\ee
%%%%%4%%10
The modified field strength $F_{\mu\nu}$ in (8) is no longer invariant under the gauge transformations (9) and (10),
\be
  F_{\mu\nu}(x) \to  F'_{\mu\nu}(x)=F_{\mu\nu}(x) + g\left(\phi_{\mu\ld} \p^\ld \p_\nu \Ld(x) -\phi_{\nu\ld}\p^\ld \p^\mu \Ld(x)\right)
\ee
%%%%%%5%%11
$$ 
 +g\left([\p_\mu \p_\ld \Ld(x)] \p^\ld A_\nu - [\p_\nu \p_\ld \Ld(x)] \p^\ld A_\mu\right) \ne F_{\mu\nu}(x).
$$ 
 We stress that the violation of the electromagnetic $U_1$ symmetry in the Lagrangian (8) is due to the general principle of gauge symmetry.
It appears that no matter how $\phi_{\mu\nu}(x)$ transforms in (10), the modified electromagnetic field strength $F_{\mu\nu}$ in (8) cannot be invariant under $U_1$ gauge transformation, except in the special case $g=0$, i.e., in the absence of gravity. To be specific, if one replaces $\p_\mu \p_\nu \Ld$ in (10) by an arbitrary infinitesimal function $X_{\mu\nu}(x)$, one cannot find a solution of $X_{\mu\nu}(x)$ for $F_{\mu\nu}(x)$ in (11) to be invariant. 

We note, however, that $F_{\mu\nu}$ transforms properly under $T_4$ gauge transformations, so that the $T_4$ gauge symmetry of Yang-Mills gravity remains intact in the presence of electromagnetic interactions\cite{3,1}.  The appearance of the $T_4$ gauge covariant derivative $\De_\mu=\p_\mu +g\phi_{\mu\ld}\p^\ld$  in the electromagnetic Lagrangian $L_{em}$ in (8) is intimately related to the universal coupling of the gravitational field to all fields in nature. 

  Such a gravitational violation of the electromagnetic $U_1$ symmetry might be experimentally tested by measuring the angle of deflection of light by the sun.  To calculate a theoretical prediction, we first derive the modified eikonal equation of Maxwell's equation in the presence of Yang-Mills gravity.  The electromagnetic Lagrangian (8) leads to the modified Maxwell equations in the presence of Yang-Mills gravity,
\be
 \Delta_\mu(\Delta^{\mu} A^{\b}-\Delta^{\b}A^{\mu}) +
(\p_{\a}J^{\a}_{\mu})(\Delta^{\mu} A^{\b}-\Delta^{\b}A^{\mu}) =0,
\ee
%%%%%%%6%%12

\section{Geometric-optics limit with and without violation of $U_1$ gauge symmetry}
As usual, the limiting expression for the field $A^\mu$ in the geometric-optics limit takes the form\cite{4}
\be
A^\mu =a^{\mu}exp(i\Psi),
\ee
%%%7%%13
where the eikonal $\Psi$ and the wave vector $\p_\mu \Psi$ are very large.

The modified electromagnetic wave equation (12) in the  geometric-optics limit leads to
%$$ \hspace{0.9in} 
\be
 [\d_\mu^\b G^{\a\s} \p_\a \Psi \p_\s \Psi] a^\mu=0,  \ \ \ \ \  G^{\a\s} = J_{\ld}^\a J^{\ld\s},   \ \ \ \   (without  \ \p_\mu A^\mu=0),
\ee
%%%%%%%%%%%%%65%8.35%%%8%%14
where  the terms involving $(\p_{\a} J^\a_{\mu})$ are small and negligible, and the terms involving $\p_\a \p_\s A^\mu$ or $\p_\a \Psi \p_\s \Psi$ are large. 
If one imposes the usual gauge condition $\p_\mu A^\mu=0$, the second term in (12) can be written as
$$
- \D_\mu \D^\b \p_{\nu}A^\mu = -J^\s_\mu \p_\s(J^{\b\a} \p_\a A^\mu)=-(\d^\s_\mu + g\phi^\s_\mu)\p_\s (J^{\b\a} \p_\a A^\mu)
$$
\be
 \approx -g\phi^\s _\mu J^{\b\a}\p_\a \p_\mu A^\mu - g\phi^\s_\mu J^{\b\a}\p_\s \p_\a A^\mu =  - g\phi^\s_\mu J^{\b\a}\p_\s \p_\a A^\mu,
\ee
%%%%%%%9 %%15
Thus, the modified Maxwell's equation (12) can be written as
\be
[ \d_{\mu}^{\b} G^{\a\s} \p_\a \Psi \p_\s \Psi -gJ^{\b\s} \phi^\a_\mu \p_\a  \Psi\p_\s \Psi] a^\mu=0, \ \ \ \ \ \   (with \ \p_\mu A^\mu=0).
\ee
%%%%10%%16

 Since we are interested in the law for the propagation of light rays and the further simplification of (16), we have expressed the amplitude $a^\mu$ in terms of the space-like polarization vector $\ep^{\mu}(\ld)$, i.e., $a^\mu = \ep^\mu(\ld)b(x), \ b(x) \ne 0$ in the limiting expression for $A^\mu$.  As usual, $\sum_{\ld} \ep^{\mu}(\ld)\ep^{\nu}(\ld) \to -\e^{\mu\nu}$ by summing over all polarizations.\cite{5}
  Multiplying $ [\d_\mu^\b G^{\a\s} \p_\a \Psi \p_\s \Psi] a^\mu\equiv Z^\b_\mu a^\mu $ in (14) by $a^\nu \e_{\nu\b}/b^2$ and summing over all polarizations, we obtain 
\be
(1/b^2) \sum_{\ld}Z^\b_\mu a^\mu a^\nu \e_{\nu\b }= - b^2 \de_\b^\mu Z^\b_\mu=0.
\ee
%%%%%%6%%%%%%%%11%%17
  After some calculations, we obtain new eikonal equations with effective metric tensors $G^{\mu\nu}$,
  \be
G^{\mu\nu} \p_\mu \Psi \p_\nu \Psi  = 0, \ \ \ \ \ \    G^{\mu\nu} = J_{\ld}^\mu J^{\ld\nu},  \ \ \ \   (without  \ \p_\mu A^\mu=0);
\ee
%%%%%%%%3%%%6%%%%%%7%%%8.36%%12%%18
\be
G_L^{\mu\nu} \p_\mu \Psi \p_\nu \Psi  = 0, \ \ \ \  G_L^{\mu\nu}  = G^{\mu\nu} -\frac{g}{4}\phi_{\ld}^\mu J^{\ld\nu}, \ \ \  (with \  \p_\mu A^\mu=0).
\ee
%%%%%%%%%%%%%%4%%%%%%%7%%%%%13%%19
Thus, we have derived the Einstein-Grossmann (EG) equations (18) for the propagation of
a light ray in an inertial frame in the geometric-optics limit.
We distinguish between the two different effective metric tensors $G^{\mu\nu}$ and $G_L^{\mu\nu}$ in (18)  and (19) because the electromagnetic Lagrangian (8) is not $U_1$ gauge invariant and this electromagnetic $U_1$ non-invariance can be tested.   If Yang-Mills gravity were to preserve $U_1$ gauge symmetry, one could choose a gauge condition such as $\p_\mu A^\mu=0$.  In this case, the EG equation would involve the effective metric tensor ${G}^{\mu\nu}_L =G^{\mu\nu}-(g/4)\phi^\mu_\ld J^{\ld\nu}$, as shown in (19).

    The use of a gauge condition in (18) is for the purpose of experimental comparison between Yang-Mills gravity (with $T_4$ symmetry, which violate $U_1$ symmetry) and some other (hypothetical) formulations of gravity with a similar tensor field with similar eikonal equation and without the violation of the $U_1$ gauge symmetry.

It must be stressed that these simple effective metric tensors in the EG equation (18) are valid only in the geometric-optics limit.  In general, eikonal equations with non-zero masses have a very complicated dependence on the wave vector $\p_\mu \Psi$, making it difficult to identify the effective Riemannian metric tensors in the EG equation, if the geometric-optics limit is not taken.  

In order to calculate the deflection of a light ray near the sun, we follow the usual convention and assume that the motion of a light ray
is in the plane defined by $\theta =
\pi/2$ in the spherical coordinates $ (t, \rh, \theta, 
\phi)$ (the x-y plane in Cartesian coordinates in an inertial frame).  Let us first consider the case $G^{\mu\nu}$ in (18), which is obtained without imposing the condition $\p_\mu A^\mu=0$.  The effective metric tensor $G^{\mu\nu}$ of the EG equation is the same as that of the perihelion shift.\cite{1}  The EG equation (18) can be written as
\be
G^{00}\left(\frac{\p \Psi}{\p t} \right)^2 + G^{11}\left(\frac{\p \Psi}
{\p \rho}\right)^2 - \frac{1}{\rho^2} \left(\frac{\p \Psi}{\p \phi}\right)^2 = 0,
\ee
%%%%%%%%%%%%%%%%68%%47%%%%54%%%9.23%%14%%20
$$
G^{00} = 1 +\frac{2Gm}{\rh}, \ \ \    G^{11}= -1+ \frac{2Gm}{\rh},    \ \ \ \ (without \ \p_\mu A^\mu=0),
$$
where $G^{33}=-1/\rh^2$, $g^2=8\pi G$ and  $m$ is the mass of the sun.  Similar to the general procedure for solving the Hamilton-Jacobi equation in a spherically symmetric tensorÊ
field, we look for the eikonal $\Psi$ in the form\cite{4,3}
\be
\Psi = -\om_0 t + M\phi + f(\rho),
\ee
%%%%%%%%%%%%%%48%%%%55%%%%9.24%%%15%%21
with constant energy $\om_o$ and angular momentum $M$ of the light ray under consideration.  One can then determine $f(\rho)$ and solve  for the trajectory of
the ray determined by the equation $\p \Psi/\p M $=constant.\cite{4}  We have
\be
\frac{d^2 \sigma}{d\phi^2} = - \sigma (1+ Q_{o}) + 3Gm \sigma^2, \ \ 
\  \sigma=\frac{1}{\rho}
\ee
%%%%%%%%%%%%%%%%%%%%%%57%%%9.26%%%16%%22
where  the  correction term $Q_{o}$ is of the order of  $G^{2}m^{2}/R^2 $, i.e., $10^{-12}$, which is extremely small and negligible.   Following the usual procedure for calculating the deflection of light due to the sun,\cite{4} we find the following results when the deflection angle is measured in an inertial frame,
\be
\Delta \phi \approx \frac{4Gm}{ R}  \approx 1.75^{\prime \prime},  \ \       (without \  \p_\mu A^\mu=0).
\ee
%%%%%17%%23
It is the deflection of a light ray passing through the spherically symmetric
tensor field generated by the sun at a distance $R=M/{\omega_o}$ from the center of the sun
to the first order approximation.  We stress that the electromagnetic Lagrangian $L_{em}$ in (8) and Yang-Mills gravity based on the replacement (1) are formulated in inertial frames in which space and time coordinates and angles have well-defined operational meaning.

For an experimental test of the possible violation of the electromagnetic $U_1$ gauge symmetry, we now consider the consequences of (19) with the gauge condition $\p_\mu A^\mu=0$.   The effective metric tensor $G_L^{\mu\nu}$ of the EG equation is similar to (18), but it is slightly different,\cite{2} 
\be
G_L^{00}\left(\frac{\p \Psi}{\p t} \right)^2 + G_L^{11}\left(\frac{\p \Psi}
{\p \rho}\right)^2 - \frac{1}{\rho^2} \left(\frac{\p \Psi}{\p \phi}\right)^2 = 0,
\ee
%%%%%%%%%%%%%%%%68%%47%%%%54%%%9.23%%18%%24
$$
G_L^{00} = 1 +\frac{7Gm}{4\rh}, \ \ \     G_L^{11}= -1+ \frac{7Gm}{4\rh},  \   \ \ \   (with \ \p_\mu A^\mu=0),
$$
and $G_L^{33}=-1/\rh^2$.  Following the same calculational steps, we obtain
\be
\frac{d^2 \sigma}{d\phi^2} = - \sigma + \frac{21 Gm}{8} \sigma^2,
\ee
%%%%%%%%19%%25
which is slightly different from (20). In this case, the deflection angle measured in an inertial frame is given by
\be
\Delta \phi \approx \frac{7Gm}{2 R}  \approx 1.53^{\prime \prime}, \ \ \ \  (with  \   \p_\mu A^\mu=0),
\ee
%%%%%%%%%%%%%%%%%%A13%%%58%%%%20%%26(26)

 Because the accuracy of measurements of the deflection of light (at optical frequencies) by the sun have been no better than (10 $-$ 20)\%,\cite{7}  both results (23) and (25) are consistent with experimental observations of $\Delta \phi_{exp} \approx 1.75^{\prime \prime}$.\cite{4}

\section{Discussion}
As usual, if the electromagnetic $U_1$ gauge symmetry is not violated by Yang-Mills gravity, one can impose the gauge condition $\p_\mu A^\mu=0$ to predict the angle of deflection as $\Delta \phi  \approx 1.53^{\prime \prime}$.  The fact that we have two different results in (23) and (26) indicates that electrodynamics is not gauge invariant in the presence of  Yang-Mills gravity.  Furthermore, quantum Yang-Mills gravity predicts that the electric charge of the electron by itself is not absolutely conserved in the presence of gravity.  The departure of the electron charge conservation in QED due to the presence of Yang-Mills gravity is extremely small and cannot be detected by present experiments.\cite{6} 

 We specifically discuss the violation of the electromagnetic $U_1$ symmetry by the Yang-Mills gravity, which is characterized by an extremely small gravitational coupling constant.   Since Yang-Mills gravity is formulated in inertial frames,  we have derived rules for Feynman diagrams and calculated explicitly the self energy of graviton and other Feynman amplitudes.\cite{3}  There is no violation of unitarity because the $T_4$ gauge invariant Lagrangian satisfy all the general requirement of a quantum field theory.  In general, all corrections related to the S matrix are extremely small due to the small gravitational coupling constant.  Therefore, there are no significant phenomenological implications within electromagnetism or within quantum electrodynamics due to the violation of the $U_1$ symmetry by the Yang-Mills gravity with the extremely small size of the gravitational coupling constant.

One may have $
F_{\mu\nu} = J_{\mu}^{\alpha} J_{\nu}^{\beta} (\partial_{\alpha} A_{\beta} - \partial_{\beta} A_{\alpha})$, which is $U_1$ gauge invariant if one assumes that the tensor field $\phi_{\mu\nu}$ does not change under the $U_1$ gauge transformation.  However, it has little to do with Yang-Mills gravity. The reason is that in the formulation of Yang-Mills gravity, once
 we assumed it to be a gauge field theory with the local space-time translation ($T_4$) symmetry, we must have the $T_4$ gauge covariant derivative to replace the usual partial derivative in the usual electromagnetic Lagrangian.  Thus we have the modified electromagnetic Lagrangian  (8), where the $T_4$ gauge covariant derivative   $( \p _\mu  -ig\phi_{\mu\nu} p^\nu) \equiv J_{\mu\nu}\p^\nu$  dictates the interactions between the gravitational field and the electromagnetic fields. 
 
 The identification of the physical meaning of $\om_0$ and $M$ are generally used  and accepted in solving the Hamilton-Jacobi type equation,\cite{4} as shown in equation (15) for the  eikonal.  Since the eikonal $\Phi$ is related to $A_\mu$ only through equation (7) when $\Phi$ and $\p_\mu \Phi$ are very large, it does not appear that these constants equal to appropriate derivatives of the Lagrangian (8).

Comparisons of results from Yang-Mills gravity and those from general relativity should be made with caution because the calculations in Yang-Mills gravity are carried out in inertial frames, while the corresponding results in general relativity are not calculated in inertial frames.  The operational meaning of coordinates and momentum in general relativity appears to be highly non-trivial.\cite{8}  Our claim is based on the discussion by Wigner, who wrote: `The basic premise of this theory [the general theory of relativity] is that coordinates are only auxiliary quantities which can be given arbitrary values for every event.  Hence, the measurement of position, that is, of the space coordinates, is certainly not a significant measurement if the postulates of the general theory are adopted: the coordinates can be given any value one wants.  The same holds for momenta.  Most of us have struggled with the problem of how, under these premises, the general theory of relativity can make meaningful statements and predictions at all'.\footnote{ Wigner said:``Evidently, the usual statements about future positions of particles, as specified by their coordinates, are not meaningful statements in general relativity.  This is a point which cannot be emphasized strongly enough...   Expressing our results in terms of the values of coordinates became a habit with us to such  a degree that we adhere to this habit also in general relativity, where values of coordinates are not per se meaningful."\cite{8}  }

If  the accuracy of the measurements of the deflection of light rays (at optical frequencies) by the sun can be improved to the level of a few percent, then one can distinguish between the results in (23) and (26).\footnote{Nevertheless, one may argue on the basis of dimensional considerations that the error from using the geometric optics limit is of the order of (wavelength)/(characteristic system size) and is exceedingly small for the deflection of light experiments with radio frequencies.  The total data for the deflection of light experiment appears to support the result (23).\cite{7,9,10,11,12,13}  
Thus,  these experiments suggest that  the electromagnetic $U_1$ gauge symmetry is violated by Yang-Mills gravity\cite{7,14}.}  The significance of such an experiment cannot be over emphasized since it would also determine whether electric charge is absolutely conserved.  According to quantum Yang-Mills gravity, the universal coupling of the gravitational field to all fields in nature implies that all internal gauge symmetries will have very small non-invariances in the presence of gravity.  In other words, all internal charges such as the color charges of quarks, the baryon charge (i.e., baryon number) and the lepton charge (i.e., lepton number) are not absolutely conserved by themselves in the presence of gravity.  In the future, such experiments of charge conservations could determine whether the Yang-Mills idea of gauge symmetry for all interactions is consistent with experiments. 

We thank reviewers for their useful comments.

%\newpage

\bibliographystyle{unsrt}

\end{document}